# Absolute Reference Energy to Realign the Band-edges of Inorganic Semiconductors Using First-principles Calculations

Tilak Das†, Xavier Rocquefelte‡*, and Stèphane Jobic*

*Institut des Matériaux Jean Rouxel, UMR 6502 CNRS – Université de Nantes, Boîte Postale 32229, 44322 NANTES Cedex 3, France.*

† *Current Address: Dipartimento di Scienza dei Materiali, Universita' degli Studi Milano-Bicocca, Via Roberto Cozzi, 55, Milano, 20125, Italy; Email: tilak.das@unimib.it*

‡ *Current Address: Institut des Sciences Chimiques de Rennes UMR 6226, Université de Rennes 1, Campus de Beaulieu, 35042 Rennes, France; Email: xavier.rocquefelte@univ-rennes1.fr*

**Abstract**

The challenge of finding an *absolute* reference energy from first-principles simulations to realigning semiconductor's valence band-top and conduction band-bottom, a theoretical methodology is proposed based on plane-wave calculations as implemented within state-of-art density functional theory. We have studied some of inorganic binary semiconductors, including both oxides and non-oxides, as for example rutile- and anatase $TiO_2$, wurtzite ZnO, rutile $SnO_2$, blende phase of GaP, GaAs, InP, ZnTe, CdS, CdSe, and SiC, those are well known and qualitatively important for photo-electrochemical, optoelectronic device applications in their standalone and/or hetero-structure morphologies. The calculated band-edges of these well known semiconductors are realigned with respect to our proposed *absolute* vacuum reference energy, which is defined with our proposed corrections and compared to their available experimental values from flat-band measurement. The prediction is reasonably well agreed with known experimental flat-band measured data. Our estimated mean absolute error bar for these set of eleven compounds is ~ 0.17 eV, closer to the known experimental limit 0.10-0.20 eV.

## 1. Introduction

Role of redox level realignment of semiconductor or insulators have immense importance in physics, chemistry and electrochemistry.[1] The hetero-junction of abrupt semiconductors used for the solid state device making due to their charge carrier transport across and along the interface led us to the devices applications as like photodiodes, LEDs, MOSFETs, MESFETs etc. Thus, knowledge on valence and conduction band edges discontinuity induced energy-offsets is quite crucial.[2-3] Also in energy scenario, concerning the current scenario of downing of fossil fuels and environmental pollution issues, the electrolysis of water or organic hydrocarbons, fuel cell technology for clean and renewable energy capturing from solar energy via the photo sensitized electrolysis or heterogeneous electrolysis/electro-catalysis (PEC) have got very much attention now as days.[4-8] In particular, the PEC reactions (both cathodic and anodic), is controlled by many aspects and one of those key quantities, is the redox level realignment of reactants upon hetero-junction formation between redox couples which has become subject of current interest in electrochemistry.[9-13] Thus, developing a methodology from theoretical approach to fulfill this requirement with reasonable accuracy might help to solve the redox energy-level predictions in molecules or band-edges realignment in a cost effective manner.

In experimental context, the redox levels of materials involved with PCE reactions is commonly measured though the flat-band measurements for standalone electrode materials.[14-15] Different theoretical approaches came forward to account for proper band-edges lines-up *i.e.* slab-vacuum approach[16], method based on the two-solids interface model[17], valence-band offset technique based on transition level of hydrogen in two-solids hetero-junction model[19], solid-liquid interface model[20] aided with first-principles calculations and/or molecular dynamics simulations. All these methods are basically, concerned with finding of an intrinsic quantity of the modeled materials from their calculations, which can be used for the line-up of band-edges in a same energy scale. In the recent past, Greiner *et al. 2011* have proposed a model experimental strategy using ultraviolet photo-luminescence measurement which uses the work-function and ionization potential energy band-offset for inorganic and organic semiconductor's non-reactive interface to align the band-edges[12]. Authors have validated that the Fermi-level-pinning transition is a universal phenomena from their *in-situ*-prepared large numbers of transition metal oxides and hence universal band-edge realignment. Thus, a reliable understanding of the semiconductor's valence- and conduction-band edge's discontinuity and alignment will allow one to tuning their device efficiency and stability. This is indeed the main motivation of the present manuscript which is devoted to find an *absolute* energy reference to realign the band-edge energies of some known inorganic semiconductors either using Standard Hydrogen Electrode (SHE) or Vacuum Energy Scale from first-principles density functional theory calculations.

The limitation of density functional theory (DFT) based first-principles calculation even with excellent DFT energy functional or potential choice, would not allow one to calculate *absolute* vacuum energy level, since it dependents on material's properties and surface termination of the slab-model[16] used in theory, involved with periodic boundary conditions.[21] Motivated by proposed free atom energy reference level model of the *Harrison 1987*, we have proposed the correction to the slab-vacuum model invented by *Frensley and Kroemer, 1977*.[16,21] Indeed, we have used an external probe atom (*i.e.* an inert isolated atom) at the center of the vacuum in a slab-vacuum model which can detect the effect of the material properties and surface effect at the *vacuum* level of the material due to long range Coulomb interaction. Then, the *intrinsic* quantity is defined as the difference of the energy level of inert atom and calculated vacuum level for different slab-vacuum models, and hence the absolute value of *vacuum zero* is estimated which is used as *absolute* reference energy for prediction.

The accuracy of our prediction was further improved by using post-DFT screened hybrid calculations using Heyd-Scuseria-Ernzerhof hybrid functional[22-24] as implemented VASP code[34] considering Hartree-Fock exact-exchange ($\alpha$) amount. The correction leads to a reasonable accuracy of the prediction compared to their available flab-band measured data for known standalone semiconductors rutile and anatase $TiO_2$, wurtzite ZnO, rutile $SnO_2$, blende SiC, GaP, GaAs, InP, ZnTe, CdS and CdSe. The estimated mean absolute error is ~ 0.17 eV which is closer to the experimental know value ~ 0.10-0.20 eV, till date.

## 2. Methodology

The well conventional slab-vacuum approach as in the periodic plane-wave DFT codes formulation is used herein.[16] The material slab/layer was grown along the non-polar surface of the each, which is one major approximation of the current approach, however. An isolated free Helium (He) atom was placed at the centre of the vacuum of each models, and hence the so called as He-slab model is prepared (see Supporting Information Figure S1 and Table S1). The implication of the total atomic local potential (sum of ionic, Hartree and exchange-correlation potential) in a periodic solid (slab model) has been elucidated elsewhere[17], which is one main guiding factor of our approach. So, the effect of the DFT exact exchange was first check for the each model. In this step the total atomic local potential was calculated on each He-slab models within 20Å vacuum He-Slab models (see Supporting Information Figure S2 for blende phase of III-V semiconductor GaP) with different DFT functionals.

Vacuum level could be one intrinsic parameter of materials to be used as reference energy level which can be used in line-up bands, as also discussed by *Van Vechten, 1985*.[26] Thus, the width of the vacuum layer is next crucial parameter to be checked. In this case, we have plotted the energy difference of the He($1s^2$) vs. corresponding vacuum level ($\Delta E_{vac\text{-}He}$) of He-slab models, for different vacuum width starting from 8-100Å thickness. See for example blende-GaP case in Supporting Information Figure S3. Because of the limitation of chosen plane-wave basis calculations with limited supercell size with vacuum 20Å, an amount of error ~ 0.1 eV comes into account in calculated vacuum label of GaP model, in order to compromise between the computation time and accuracy. So, using the vacuum width of 20Å for all He-Slab models, the accounted error was eliminated at the final predictions of *absolute* vacuum for all eleven materials studied here.

Finally, the width of the material layer was checked based on the results from the two different choices of layer thickness i.e. 7 and 15 layers of material within a He-slab model with vacuum length of 20Å (Supporting Information Figure S4). The estimated energy difference of the planar average of layer's total local atomic potential to the corresponding vacuum level ($\Delta E_{vac\text{-}layer}$) is shown in this figure. We have noted that even for the ionic type oxides rutile $TiO_2$ the 7-layers thick (~10Å) is sufficient for achieving reasonable accuracy, as the value of $\Delta E_{vac\text{-}layer}$ is same for these two models. So, we finalize for all other cases that a minimum 10Å thickness is essential to construct the He-slab model.

## *2.1. Modified Vacuum Approach*

The band-edges realignment from modified vacuum approach is illustrated in detail, considering the example case rutile $TiO_2$, as shown in Figure 1. In the top two panels [**a**] and **[b]** of this Figure 1, respectively the total DOS of bulk and layer-vacuum model are shown. The $V_{be}$ is taken to be $V_{be}(bulk) = E_F(bulk)$ and $V_{be}(slab) = E_F(slab)$, in both cases, where $E_F$ is corresponding Fermi level, with respect to their reference energy scale in bulk, $E_{Ref}(bulk)$ and slab, $E_{Ref}(slab)$ using a plane-wave first-principles calculations. Surface termination and hence polarity in a slab-model, led to an important role for defining the Fermi level, thus defining it as $V_{be}$ could not be a proper,[31] that can be use for alignment. Since, we have used a sufficiently thicker slab material; we can imagine that the bulk of the slab is equivalent to the pure bulk phase of the material. In some extend, the average core potential of bulk atoms of these two systems will same as we have kept fix the experimental volume unchanged in slab model. Hence, we have calculated the core level shift, using Eq. 1 as below given:

$$\Delta E_{core} = \{E_{core}(slab) - E_{core}(bulk)\}\ldots\ldots\ldots(1).$$

Inserting the core level shift, $\Delta E_{core}$ over the calculated DOS of bulk phase, one can estimate the surface pollution free Fermi level ($V_{be}$) from a slab model, within a same reference energy i.e. $E_{Ref}(slab)$, as shown in the panel [**c**] of the **Figure 1**. This superposition procedure is quite reasonable way than the Fermi level shift, because from our estimation in case of rutile $TiO_2$, accounted error due to surface effects is ~ 0.24 eV, which was excluded from the estimation of $V_{be}$. Finally, to align the $V_{be}$ we have used the vacuum energy, $E_{vacuum}(slab)$ calculated from the slab model. Thus, the vacuum-level shift is considered with following Eq. 2

$$\Delta E_{vacuum} = \{E_{vacuum}(absolute\ ) - E_{vacuum}(slab)\} = - E_{vacuum}(slab) \ldots\ldots\ldots (2).$$

Because, the absolute vacuum energy is 0.00 eV (cf. Figure 1[**d**]). In a similar strategy, the $V_{be}$ is predicted for rest of the other slab-models, and the so called "Modified Vacuum Approach" is proposed herein.

## *2.2. Helium (He) Atom in the "Modified Vacuum Approach"*

The vacuum level, $E_{\text{vacuum}}$ calculated considering the ionic, Hartree and exchange-correlation potential, is actually affected due to the long-range Coulomb interaction between layers in a periodic boundary condition of the chosen potential, spherical atomic radii using a DFT tool. So, calculated vacuum level is *pseudo* like. Hence, considering a finite size vacuum length along the surface grown direction, led to an error in the estimated vacuum energy level. This obviously tells a necessity of probe that can help to eliminate the accounted error in the calculated vacuum level. The choice of such probe is the free Helium (He) atom, a noble gas element. The probe atom should be such that, 1) it need to be smaller in size, which allows us to use smaller supercell size, 2) the chemical inertness of the probe would help not to have strong interaction with neighboring layers or itself, 3) the active energy level of the probe should be far below in the energy scale than the chemical reaction/hybridization energy range of mostly used semiconductors or redox level of molecular species. The difference of the current approach than the previously mentioned “Modified-Vacuum Approach” is that now we have used the He($1s^2$) of the probe as reference energy for correcting the vacuum energy level and hence the so called “He-Slab model” is proposed.

For an free He atom, experimentally the ionization energy is +24.59 eV i.e. the He($1s^2$) level locates at the -24.59 eV below the Fermi level, which is quite lower than presently used semiconductor’s valence band top-edges. So, the correct description of the He($1s^2$) energy level from ground state DFT is matter of question, which should be figured out. In all the He-slab models the average value calculated He($1s^2$) level, using PBE-GGA functional (Hartree-Fock exact exchange $\alpha$ = 0.00), is -15.64 eV *i.e.* 36% smaller than the experimental value. In all case of He-Slab this is studied and for GaP model this $\alpha$ dependence of the He($1s^2$) energy level and the core potential shown for four different value of $\alpha$ = 0.00, 0.20, 0.25, 0.73 in the Supporting Information Figure S5. Thus, we concluded that a significant amount of the Hartree-Fock exact exchange must be added via $\alpha$ value to better describe the He($1s^2$) energy level $E_{He},$ as it is going closer to the experimental known data. Considering, a linear extrapolation up to $\alpha$ = 1.00 of the $E_{He}$ values of all He-Slab models and their calculated values at $\alpha$ = 0.00 and 0.25 using HSE functional, the average value of He($1s^2$) energy is estimated to be -23.86 eV (at $\alpha$=1.00). This is little over estimated but quite closer to the experimental one i.e. -24.59 eV (See Supporting Figure S6). Thus, it is significantly important to use a suitable value of $\alpha$ to better describe the atomic orbital energy positions, as described elsewhere.[27] We will also see in the results section that such approach using He-slab model, that how it will help us to reduce the mean absolute error or standard deviation of our prediction.

However, the exact description of the He energy level is not main purpose herein. As said previously, the He($1s^2$) plays as a probe in the model and help us to correct the calculated vacuum energy level of each semiconductor's He-Slab models. Such estimation is possible by considering the cubic box containing one He atom within a plane-wave periodic potential calculation in DFT. In the Supporting Information Figure S7, the evolution of the He($1s^2$) energy with the cell size i.e. the He-He distance (*d*), is shown for PBE-GGA calculated data (α = 0.00) or HSE functional (α = 0.25) calculation. As we can see that the convergence was reached for the cell dimension (*d*) greater than *d*=30Å and the calculated He($1s^2$) energy level is at -15.71 eV. But, from the previous He-slab model this value was estimated at -15.60 eV for He-Slab GaP model from PBE-GGA calculations. Thus, an amount of -0.11 eV is essential to correct the He energy level of the He-slab model, based on the He-box correction.

Since, all He in different He-slab models do not have same neighbor, the error will be different. Within the PBE-GGA calculated data, using Lorentz fit the error amount was evaluated for all cases and checked that these values are different as shown in the Supporting Information Table S2. It should be noted that this amount is small and shifted in the same direction.

## 3. *Results and Discussions*

### 3.1. *Modified Vacuum Approach*

We have done prediction based on the nine available experimental data of valence band top-edge, $V_{be}$ out of eleven chosen binary systems. We have shown in the **Table 1** the numerically calculated $V_{be}$ data following the Eq. 1 and Eq. 2, predicted using the theoretical formulations as given in **Figure 1**, beside the experimental data against SHE or vacuum scale reference energy scale. Our prediction for $V_{be}$ were done from the HSE06 calculations using α = 0.00 and 0.25 for Hartree-Fock exact exchange. In the PBE-GGA (α = 0.00) level, data is far from the experimental values with mean absolute error (MAE) nearly 1.5 eV and standard deviation (SD) 0.94 eV (cf. Table 1). With the increasing value of α = 0.25 using the HSE06 functional, these two error bars reduced to 0.9 and 0.7 eV, respectively. Thus, we have considered all the linear variations of the quantities needed to estimate the theoretical value of $V_{be}$ (cf. Supporting Information Figure S5) as an input from the PBE-GGA and HSE06 calculated data, using the least square fit procedure on nine equations (for nine compounds) and one degree of freedom (α value). We have obtained the best agreement at α = 0.635 from this "Modified Vacuum Approach". The graphical representation of these data using α = 0.635 is shown in the **Figure 2**. Out of all these cases, we have seen that the $V_{be}$ is moving towards the experimental data at α = 0.635, which give us the best fit with the MAE and SD respectively 0.30 and 0.44 eV (cf. Table 1).

The solid black line passes through the diagonal of the plot is the measure of accuracy of our predictions compared to the experimental data. Predicted data just falls around it and the experimental and theoretical error bar are shown for each modified slab-model prediction. The maximum overestimation for the $V_{be}$ was found for SiC compound, about +1.06 eV and underestimation about -0.37 eV was found for the rutile $TiO_2$, at α value 0.635. In our observation, this could be the impact of the He-He interaction as noted due to finite vacuum dimension and in-plane lattice size choice, which led to such larger error for SiC and R-$TiO_2$ (cf. Supporting Information Figure S7 and Table S2).

## 3.2. He-Slab Approach

Use of higher α value within the "Modified Vacuum Approach" is obviously, not enough since MEA or SD is quite larger than the experimental error bar limit 0.10-0.20 eV. Thus, incorporation of the He-He interaction is very much crucial here i.e. inclusion of He-correction to remove the pseudo behavior of vacuum and using corrected reference scale, we have estimated the values of $V_{be}$ of all eleven compounds within this approach. The presently predicted numerical values are shown in the **Table 2**. The error amount is reduced, with estimated MAE is nearly 0.14 eV and SD is 0.17 eV (cf. Table 2), those are in excellent agreement with the available experimental data. The plotting of the predicted $V_{be}$ extrapolated at $\alpha = 0.73$, against the experimental data is shown in the **Figure 3**. The maximum overestimation is found for the SiC compound and underestimation is noted for the GaP, nearly +0.16 and -0.27eV, respectively.

So, we have corrected strategy from first-principles calculation the slab-vacuum approach that can predict the $V_{be}$ for semiconductors in an absolute energy scale i.e. so called the *absolute* vacuum energy label. Now, in order to complete this prediction procedure for both the valence band-edge $V_{be}$ and conduction band-edge $C_{be}$, at the same scale, we have used two possible approaches. Either, one can add the known experimental band gap to the predicted $V_{be}$ values or use some good DFT functional to have correct band gap of the test compound that will be added with the predicted $V_{be}$ to estimate the $C_{be}$. In the **Figure 4**, both of these two band edges are shown with respect to the vacuum energy scale, where experimental band-gap is used to predict $C_{be}$. Quite reasonable agreement is found for the all calculated $V_{be}$ and hence $C_{be}$ for all nine compounds compared to their flat-band measured experimental data. No suitable experimental data for pristine anatase $TiO_2$ or blende ZnTe is found, thus experimental data of them is absent in the plot.

## 4. Outlook and Electrochemical Implication

Looking at **Figure 4**, we can clearly see one qualitative trend is obviously good for all these cases, as compared to their experimental data from *Nozik*, 1978[9] or *Grätzel*, 2001.[14] Specially, if we look to the general trend of the $C_{be}$, for rutile $SnO_2$, $TiO_2$ and wurtzite ZnO, very much systematic alignment is reproduced from our computation experiment based predictions, whereas $C_{be}$ is moved little higher in energy, respectively. Also, for other non-oxides i.e. CdS or CdSe and InP or GaAs pairs, predicted $C_{be}$ have also well agreed with the experimental data. More specifically, the prediction of relative $C_{be}$ location for two similar oxides anatase and rutile $TiO_2$ are also, correctly reproduced. Scanlon and co-workers, 2013 have proved the relative band-edge alignment of these systems rigorously from the experimental XPS (Madelung potential) and Quantum-mechanical combined Molecular-mechanical embedded DFT computations which proves relative $V_{be}$ offset ~ 0.5 eV higher in energy for rutile than anatase.[33] Indeed, considering the experimental band-gap 3.03 and 3.20 eV respectively for them, it led to ~ 0.2-0.3 eV higher energy offset of $C_{be}$ for rutile, which is similar than observed data from our prediction and validate robustness of our method.

A computational cost effective output from such theoretical methodology could help the community those are working on the photoelectrochemical, electrolysis and optoelectronics domain either experimental or theoretical. By inventing such corrected vacuum energy as an absolute scale in similar footling like experimental one, it can help for further development of redox reaction based energy storage and conversion technologies in a cost effective and efficient manner. It should be noted that the current methodology not necessarily takes into account the actual description of the real surface. Truncation in any orientation at any position of the material is sufficient, even though it is not an energetically stable surface, since we did not take into account the surface relaxations, reconstruction and combined impact of the nanoparticulate facets effect. Indeed, works from Nørskov and co-works on the pristine oxides, nitrides and sulfides for electrochemical reaction for Hydrogen Evolution Reaction (HER) and O(Oxygen)ER in such simplistic approach of DFT where, all complexity from experimental set up might not be essential to reproduce qualitative explanations of the experimental observation.[28-30]

A detail review on the electrochemical studies using DFT tool and its limitation for application towards the HER (or OER) in electrocatalysis or photocatalysis process, is discussed by Qiao and co-workers.[11] It is obvious that the current development of the DFT tool and computational power is always one crucial factor for further improvement of the models in first-principles based theoretical calculations. Thus, whether it is the periodic solid surface as shown by Neugebauer and co-workers[31] on the polar oxide surface of ZnO(0001) or the $CO_2$ reduction on a cubanes like finite cluster of Fe-Ni-S as shown by Nørskov and co-workers[32], both has demonstrated the impact of solavation on material's surface (electrode) in an electrochemical environment intact with liquid or vacuum, the reconstruction of surface takes place.

Hence, we embark our conclusion with caution that in our present approach even though we did not take into account in our computations all correct experimental variables, including the temperature to get the accurate description of the H* or OH*-mediated radicals or hydrocarbon adsorption or reduction which are controlled by the redox of the electrodes, are reproduces in a simplistic manner using our He-slab model approach. It is will be matter of future research interest for improvement of our current approach. Thus, in this scenario, not an exact description is needed rather reasonable trend of bands line-up is possible compared to the experimental flat-band measured data of highest and lowest molecular orbital in solids with a moderate error bar.

## 5. Conclusions

In conclusion, the present a strategy which is very much promising in our opinion as it can predict the $V_{be}$ for set of different oxides and non-oxides materials within a reasonable accuracy from low cost and cheap computational approach. In the He-Slab approach, both the He atom and the material feels the pseudo vacuum, thus the He($1s^2$) energy level helps to find the *absolute* vacuum level, validating the state-of-art of first-principles calculations. This is in fact the main reason why we have reached to such excellent accuracy of the prediction, with the SD ~ 0.2 eV. Indeed, our ultimate goal to find absolute reference energy scale for all solid state compound, which is qualitatively achieved. In future, we are interested to enlarge this method for the non-polar surface in order to make it application for all kind of solids and molecules. We expect such study would help to develop in the recent progresses using redox reactions based different technological applications *i.e.* electro/photo-catalysis, fuel-cell, and photovoltaics etc. solar energy conversion and storage issue in a cost effective way. Even more specifically, this tool can open up another branch of research and development in environmental friendly solar energy conversion techniques by photo-sensitization.

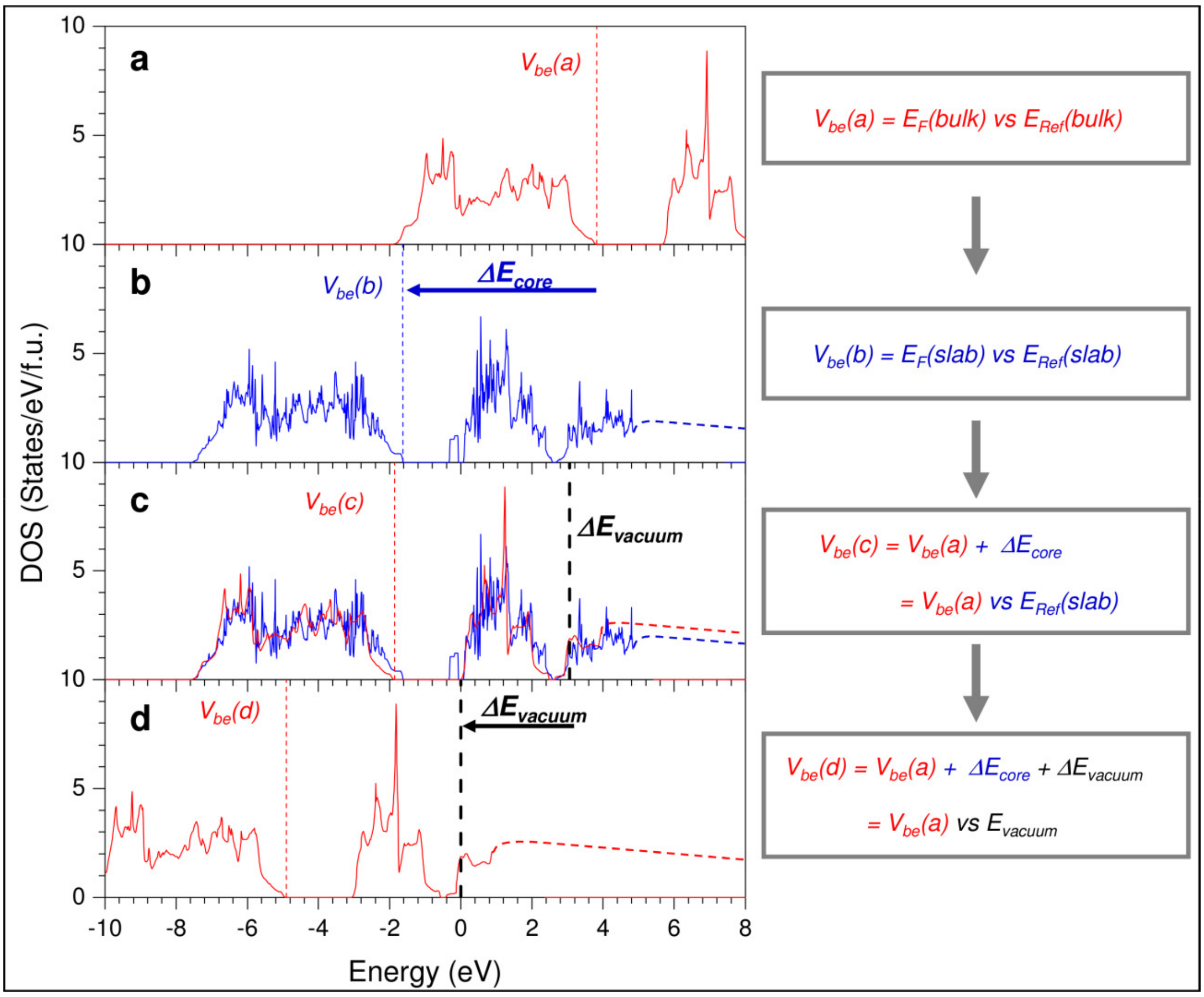

***Figure 1.*** The methodology of valence band edge $V_{be}$ prediction considering the example case R-$TiO_2$. In the panel [**a**] and [**b**], the total DOS of the bulk and slab-model is shown, respectively. $V_{be}$ is taken to be the corresponding calculated Fermi level, w.r.t. their reference energy $E_{Ref}$, respectively. In the [**c**] panel, they are superimposed based on their average core energy shift ($\Delta E_{core}$). Finally, in the panel[ **d]** the $V_{\mathrm{be}}$ is shown excluding the surface impurity level and rescaled w.r.t. vacuum level shift ($\Delta E_{vacuum}$).

***Table 1.*** The predicted valence band top-edges ($V_{be}$) of studied semiconductors using "Modified Vacuum Approach" and their comparison with the available experimental data from Flab-band measurements.

| | Experimental | | Theoretically Calculated | | | | | |
|---|---|---|---|---|---|---|---|---|
| | | | $V_{be}$ | | $V_{be}$ | | $V_{be}$ | |
| **Vacuum-Slab Model** | **$V_{be}$ (NHE) (eV)** | **$V_{be}$ (Vacuum) (eV)** | **α=0.00** | **Error ΔE (eV)** | **α=0.25** | **Error ΔE (eV)** | **α=0.635** | **Error ΔE (eV)** |
| $SnO_2$ | -3.89 | -8.39 | -5.44 | -2.95 | -6.46 | -1.93 | -8.03 | -0.36 |
| R-$TiO_2$ | -3.11 | -7.61 | -4.91 | -2.70 | -5.82 | -1.79 | -7.24 | -0.37 |
| CdS | -2.18 | -6.68 | -5.18 | -1.50 | -5.73 | -0.95 | -6.58 | -0.10 |
| A-$TiO_2$ | - | - | -5.91 | - | -6.84 | - | -8.26 | - |
| ZnO | -3.03 | -7.53 | -5.22 | -2.31 | -6.17 | -1.36 | -7.63 | 0.10 |
| SiC | -1.72 | -6.22 | -5.79 | -0.43 | -6.38 | 0.16 | -7.18 | 1.06 |
| CdSe | -1.66 | -6.16 | -5.15 | -1.01 | -5.63 | -0.53 | -6.38 | 0.22 |
| InP | -1.33 | -5.83 | -4.81 | -1.02 | -5.16 | -0.68 | -5.69 | -0.14 |
| GaP | -1.30 | -5.80 | -4.94 | -0.87 | -5.30 | -0.50 | -5.86 | 0.06 |
| GaAs | -0.84 | -5.34 | -4.79 | -0.55 | -4.93 | -0.41 | -5.16 | -0.18 |
| ZnTe | - | - | -4.81 | - | -5.25 | - | -5.93 | - |
| MAE(SD) | | | 1.48 | (0.94) | 0.92 | (0.69) | 0.29 | (0.44) |

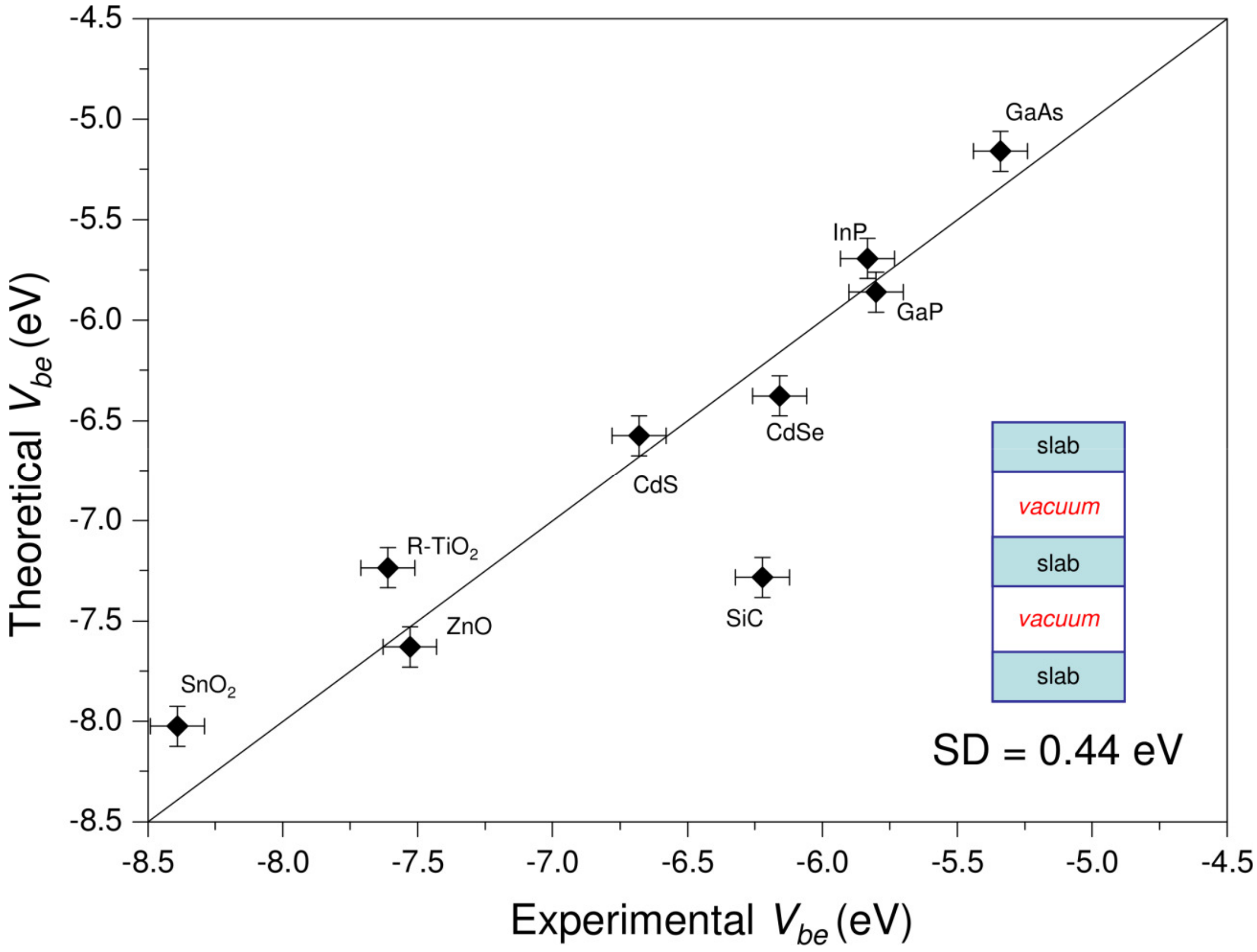

***Figure 2.*** The predicted values of $V_{be}$ using our "Modified Vacuum Approach" derived from the extrapolated data using α = 0.635 within HSE calculations.

***Table 2.*** The predicted valence band top-edges ($V_{be}$) of studied semiconductors using "He-Slab Approach" and their comparison with the available experimental data from Flab-band measurements.

| | Experimental | | Theoretically Calculated | | | | | |
|---|---|---|---|---|---|---|---|---|
| | | | $V_{be}$ | | $V_{be}$ | | $V_{be}$ | |
| **Vacuum-Slab Model** | **$V_{be}$ (NHE) (eV)** | **$V_{be}$ (Vacuum) (eV)** | **α=0.00** | **Error ΔE (eV)** | **α=0.25** | **Error ΔE (eV)** | **α=0.73** | **Error ΔE (eV)** |
| $SnO_2$ | -3.89 | -8.39 | -5.52 | 2.87 | -6.49 | 1.90 | -8.37 | 0.02 |
| R-$TiO_2$ | -3.11 | -7.61 | -4.85 | 2.76 | -5.73 | 1.88 | -7.44 | 0.17 |
| CdS | -2.18 | -6.68 | -5.15 | 1.53 | -5.68 | 1.00 | -6.70 | -0.02 |
| A-$TiO_2$ | - | - | -5.32 | - | -6.19 | - | -7.88 | - |
| ZnO | -3.03 | -7.53 | -5.24 | 2.29 | -5.97 | 1.56 | -7.37 | 0.16 |
| SiC | -1.72 | -6.22 | -4.99 | 1.23 | -5.34 | 0.88 | -6.02 | 0.21 |
| CdSe | -1.66 | -6.16 | -4.95 | 2.21 | -5.43 | 0.73 | -6.37 | -0.21 |
| InP | -1.33 | -5.83 | -4.61 | 1.22 | -5.00 | 0.83 | -5.74 | 0.09 |
| GaP | -1.30 | -5.80 | -4.75 | 1.05 | -5.20 | 0.60 | -6.07 | -0.27 |
| GaAs | -0.84 | -5.34 | -4.74 | 0.60 | -4.97 | 0.37 | -5.42 | -0.08 |
| ZnTe | - | - | -4.87 | - | -5.32 | - | -6.20 | - |
| MAE(SD) | | | 1.64 | (0.80) | 1.08 | (0.56) | 0.14 | (0.17) |

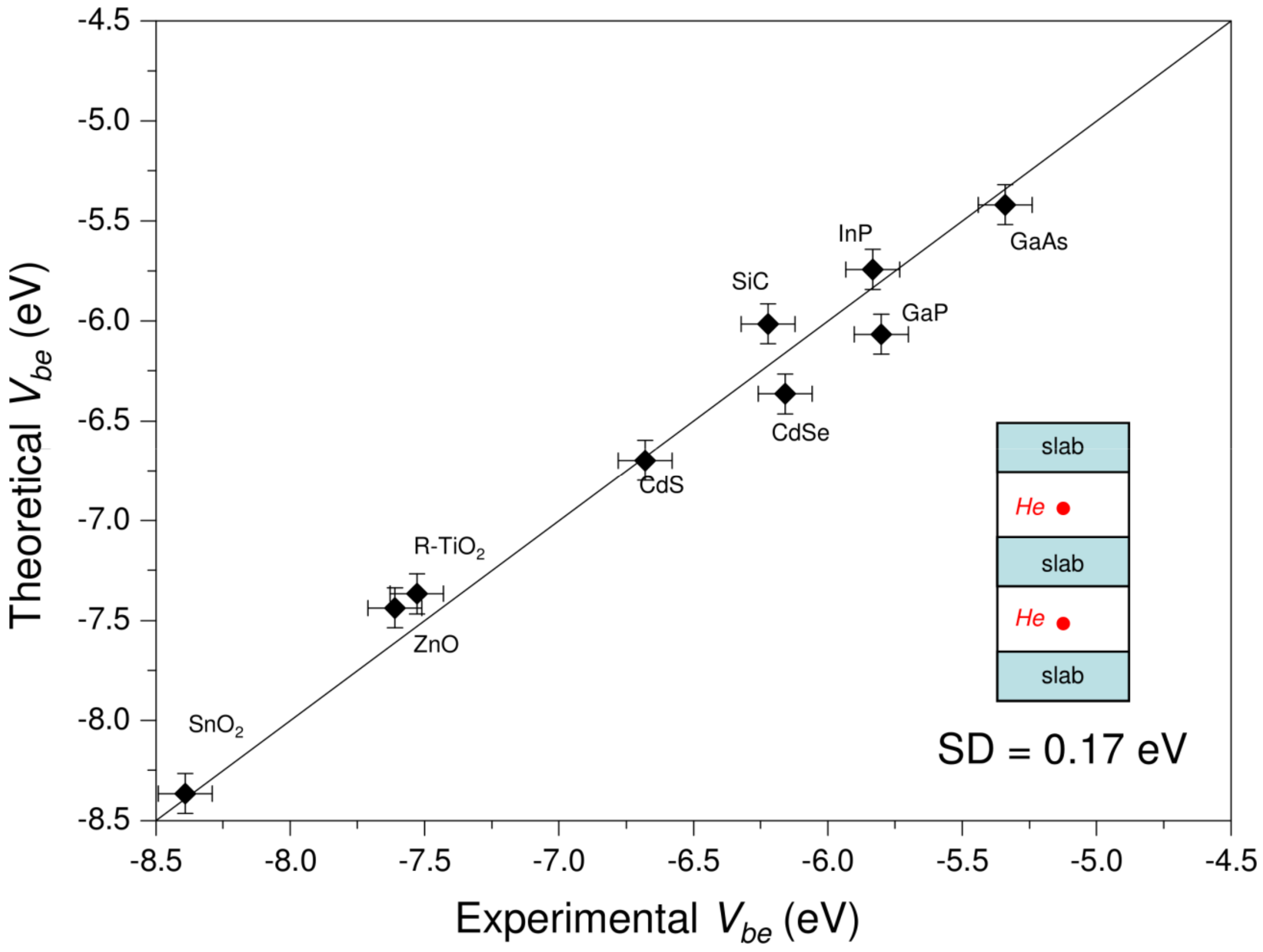

***Figure 3.*** The predicted values of $V_{be}$ using our "He-Slab Approach" derived from the extrapolated data using $\alpha = 0.73$ within HSE calculations.

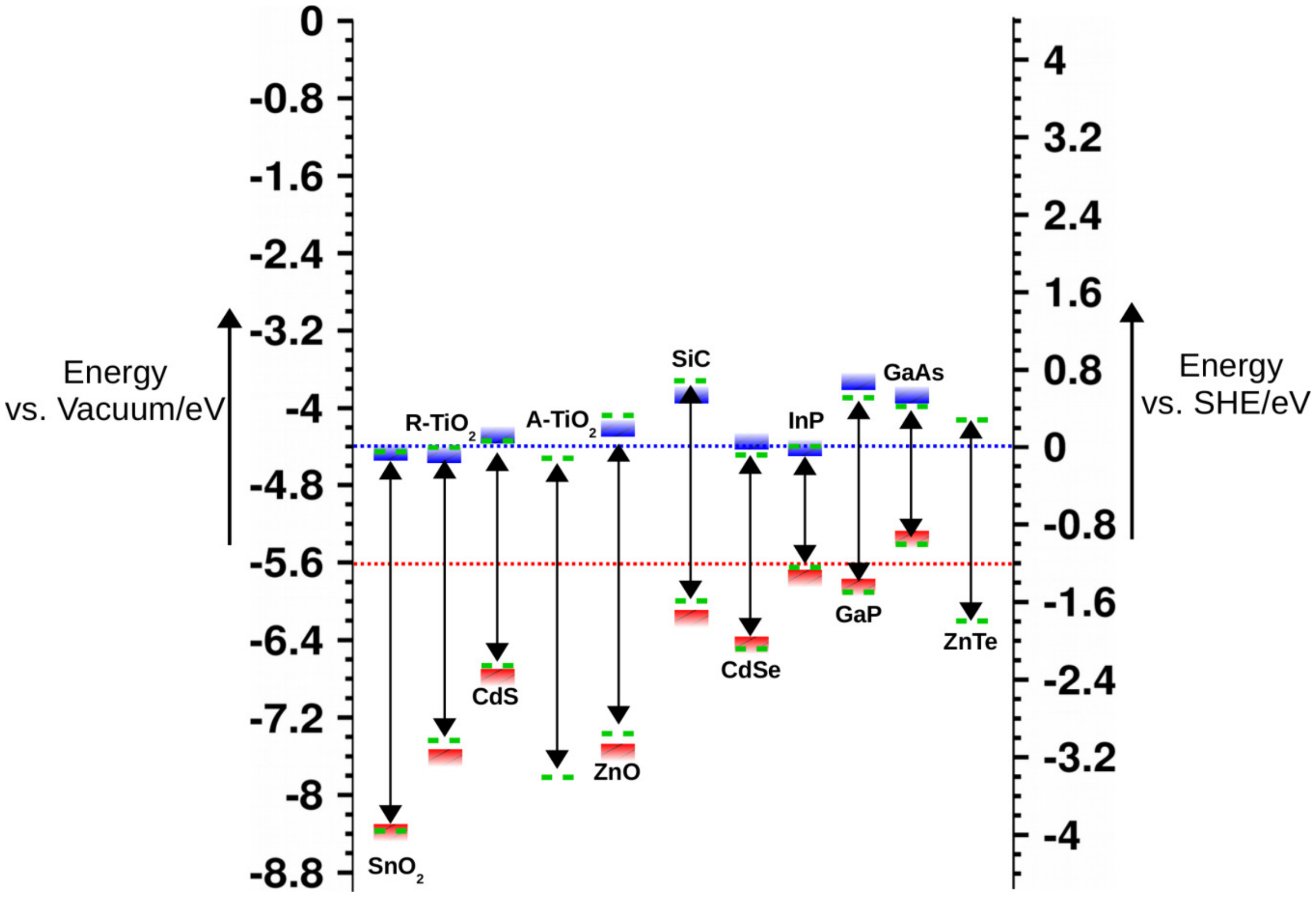

***Figure 4.*** The calculated $V_{be}$ and $C_{be}$ at α = 0.73 (green dashed bars) are represented with the available experimental data (red and blue solid bars, respectively), using vacuum energy (left) or SHE energy (right) scale. The $C_{be}$ were estimated by adding the experimental optical band-gap for each compounds with calculated $V_{be}$. The two dotted lines are the reduction (blue) and oxidation (red) redox level of pure water. Eleven materials are denoted by labeling on top or bottom of their band-edges.

## References

[1] A. Franciosi, C. G. Van de Walle, *Surf. Sci. Reports.* **1996**, *25*, 1-140.

[2] J. H. Haeni, D. G. Schlom, *MRS Bull.* **2002**, *27*, 198.

[3] C. Dean, A. F. Young, L. Wang, I. Meric, G. –H. Lee, K. Watanabe, T. Taniguchi, K. Shepard, P. Kim, P. Kim, *Solid State Comm.* **2012**, *152*, 1275-1282.

[4] F. E. Osterloh, B. A. Parkinson, *MRS BULLETIN* **2011**, *36(1)*, 17-22.

[5] K. Maeda, K. Domen, *MRS BULLETIN* **2011**, *36 (1)*, 25-31.

[6] A. Kudo, *MRS BULLETIN* **2011**, *36 (1)*, 32-38.

[7] J. Z. Zhang, *MRS BULLETIN* **2011**, *36 (1)*, 48-55.

[8] M. Kaneko and I. Okura (Eds), Photocatalysis: science and technology, Springer-Verlag Berlin Heidelberg, New York, 2002, ISBN 3-540-43473-9.

[9] A. J. Nojik, *Ann. Rev. Phys. Chem.* **1978**, *29*, 189-222.

[10] Frank E. Osterloh, *Chem. Mater.* **2008**, *20*, 35-54.

[11] Y. Zheng, Y. Jian, M. Jaroniec, S. Z. Qiao, *Angew. Chem. Int. Ed.* **2015**, *53*, 52-65.

[12] M. T. Greiner, G. G. Helandar, W. –M. Tang, Z. –B, Wang, J. Qiu, Z, -H. Lu, *Nature Materials,* **2011**, *NMAT3159.*

[13] S. Tokito, K. Noda, Y. Taga, *J. Phys. D*, **1996**, *29*, 2750-2753.

[14] M. Grätzel, *Nature* **2001**, *414*, 338-344.

[15] M. G. Walter, E. L. Warren, J. R. McKone, S. W. Boettcher, Q. Mi, E. A. Santori, N. S. Lewis, *Chem. Rev.* **2010**, *110*, 6446-6473.

[16] W. H. Frensley, H. Kroemer, *Phys. Rev. B* **1977**, *16(6)*, 2642-2652.

[17] C. G. Van de Walle, R. M. Martin, *Phys. Rev. B* **1987**, *35(15)*, 8154-8165.

[18] F. D. Angelis, S. Fantacci, and A. Selloni, *Nanotechnology* 2008, *19*, 424002.

[19] Chris G. Van de Walle, J. Neugebauer, *Nature* **2003**, *423*, 626-628.

[20] Y. Wu, M. K. Y. Chan, G. Ceder, *Phys. Rev. B* **2011**, *84*, 235301.

[21] L. Kleinman, *Phys. Rev. B* **1981**, *24*, 7412.

[22] J. Heyd, G. E. Scuseria, M. Ernzerhof, *J. Chem. Phys.* **2003**, *118*, 8207.

[23] J. Heyd, G. E. Scuseria, M. Ernzerhof, *J. Chem. Phys.* **2006**, *124*, 219906.

[24] J. Paier, M. Marsman, K. Hummer, G. Kresse, I. C. Gerber, J. G. Ángyán, *J. Chem. Phys.* **2006**, *125*, 249901.

[25] W. A. Harrison, *Electronic Structure and the Properties of Solids,* (Freeman, San Francisco, 1980), p. 253.

[26] J. A. Van Vechten, *J. Vac. Sci. Technol. B* **1985**, *3*, 1240.

[27] Y. Hinuma, A. Grünies, G. Kresse, F. Oba, *Phys. Rev. B* **2016**, *90,* 155405.

[28] J. Rossmeisl, Z.-W. Qu, H. Zhu, G.-J. Kroes, J. K. Nørskov, *J. Electroanal. Chem.* **2007**, *607*, 83-89.

[29] E. M. Fernández, P. G. Moses, A. Toftelund, H. A. Hansen, J. I. Martínjez, F. Abild-Pedersen, J. Kleis, B. Hinnemann, J. Rossmeisl, T. Bligaard, J. K. Nørskov, *Angew. Chem.* **2008**, *120,* 4761-4764.

[30] I. C. Man, H.-Y. Su, F. Calle-Vallejo, H. A. Hansen, J. I. Martínjez, N. J. Inoglu, J. Kitchin, T. F. Jaramillo, J. K. Nørskov, J. Rossmeisl, *ChemCatChem.* **2011**, *3*, 1159-1165.

[31] S.-H. Yoo, M. Tadorova, J. Neugebauer, *Phys. Rev. Lett.* **2018**, *120*, 066101.

[32] J. B. Varley, H. A. Hansen, N. L. Ammitzbøll, L. C. Grabow, A. A. Peterson, J. Rossmeisl, J. K. Nørskov, *ACS Catal.* **2013**, *3*, 2640-2643.
[33] D. O. Scanlon et al. *Nature Materials* **2013**, *12*, 798-801.
[34] G. Kresse, J. Hafner, *Phys. Rev. B* **1993**, *47*, R558; G. Kresse, J. Furthmüller, *Phys. Rev. B* **1996**, *54*, 11169.